\documentclass[reprint,
 amsmath,amssymb,
 aps,
]{revtex4-2}
\usepackage{multirow}
\usepackage{float}
\usepackage{graphicx}% Include figure files
\usepackage{dcolumn}% Align table columns on decimal point
\usepackage{bm}% bold math
\begin{document}
%\begin{CJK}{UTF8}{gbsn}
%\preprint{APS/123-QED}

\title{Quantifying Photoproduction Corrections to $J/\psi$ Measurements in Au+Au and Cu+Cu Collisions at $\sqrt{s_{\textrm{NN}}} = 200$ GeV}% Force line breaks with \\

\author{Zhejin Liu}

\author{Zebo Tang}
\email{zbtang@ustc.edu.cn (Corresponding author)}
\author{Xin Wu}
\author{Wangmei Zha}
\email{first@ustc.edu.cn (Corresponding author)}

\affiliation{
 University of Science and Technology of China, Hefei 230026, China
}
\date{\today}% It is always \today, today,
             
\begin{abstract}

The contribution of $J/\psi$ production from coherent and incoherent photon-induced production is calculated in Au+Au and Cu+Cu collisions at $\sqrt{s_{\textrm{NN}}} = 200$ GeV. The yield and nuclear modification factors ($R_\mathrm{AA}$) contributed by photoproduction are presented as functions of transverse momentum ($p_T$) and centrality at both mid- and forward rapidity. The $R_\mathrm{AA}$ of $J/\psi$ from photoproduction is found to be as high as 0.6 in the lowest $p_{T}$ bin ($0 < p_{T} < 1$ GeV/$c$) at mid-rapidity in 60-92\% Au+Au collisions. The $R_\mathrm{AA}$ measured in Au+Au and Cu+Cu collisions at $\sqrt{s_{\textrm{NN}}} = 200$ GeV are corrected by subtracting the calculated photoproduction contribution. 

\end{abstract}

%\keywords{hotoproduction}%Use showkeys class option if keyword
\maketitle

%\tableofcontents

\section{\label{sec:level1}Introduction}
$J/\psi$ suppression in heavy-ion collisions is one of the most important signatures of the existence of Quark-Gluon Plasma (QGP)~\cite{Matsui:1986dk,Brambilla:2010cs,STAR:2005gfr,Braun-Munzinger:2007edi}, due to the color screening effect. $J/\psi$s are created in initial collisions because of the large $c$ quark mass and have long enough lifetime (about 2000 fm/$c$) to pass through the medium, which would help the $J/\psi$s carry the information of the deconfined medium. Moreover, the substantial branching ratio of $J/\psi$ decaying into dileptons makes it experimentally detectable with ease. These characteristics render $J/\psi$ a valuable tool for investigating the QGP. Over the years, various experiments on $J/\psi$ production in heavy-ion collisions have been conducted. The CERN SPS NA50 have performed fixed-target experiment to study the $J/\psi$ suppression in Pb+Pb collisions ~\cite{NA50:2000brc}, which indeed found an anomalous suppression of $J/\psi$. RHIC at Brookhaven National Laboratory (USA) also conducted similar measurements but with higher energy. Two of the four experiments installed at RHIC, namely STAR and PHENIX, have measured $J/\psi$ yields in $p$+$p$, $d$+Au, Au+Au, and Cu+Cu collisions over a wide range of energies~\cite{STAR:2016uxt,STAR:2016utm,STAR:2013eve,PHENIX:2011img,STAR:2012wnc,PHENIX:2008jgc,STAR:2019fge,STAR:2009irl,STAR:2020igu,STAR:2019vkt,PHENIX:2019brm,PHENIX:2012xtg,PHENIX:2009ghc,PereiraDaCosta:2005xz}, from 39 GeV to 200 GeV per pair of nucleons. See Ref.~\cite{Tang:2020ame} for a recent review. The expectation was that $J/\psi$ suppression would be more significant compared to SPS results due to the creation of a hotter QGP. However, experimental data did not support this hypothesis and was attributed to the regeneration effect~\cite{Braun-Munzinger:2000csl,Thews:2000rj,Yan:2006ve,vanHees:2004gq}. By far, most experimental data on $J/\psi$ suppression in heavy-ion collisions can be explained as the interplay between the color-screening effect, the regeneration effect, and cold nuclear matter (CNM) effect ~\cite{Brambilla:2010cs}.  

The study of $J/\psi$ behavior in hadronic heavy-ion collisions (HHIC) was additionally stimulated by measurements made by ALICE~\cite{ALICE:2015mzu} and STAR~\cite{STAR:2019yox} at exceedingly low $p_T$. Notably, significant enhancements have been observed, signifying the existence of coherent photon-nucleus interactions in HHIC, a phenomenon previously thought to occur exclusively in ultra-peripheral collisions (UPC). Within this production mechanism, $J/\psi$ can arise from the interaction of intense electromagnetic fields accompanying relativistic heavy ions~\cite{Bertulani:2005ru}. The intense electromagnetic field can be conceptualized as a spectrum of equivalent photons using the equivalent photon approximation (EPA)~\cite{Krauss:1997vr,Vidovic:1992ik}. The quasi-real photon emitted by one nucleus can fluctuate into a $c \bar{c}$ pair, scatter off the other nucleus, and subsequently materialize as an actual $J/\psi$. Theoretical  calculations based on this production mechanism could quantitatively describe the anomalous excesses, further confirming the presence of coherent photon-nucleus reactions in HHIC~\cite{STAR:2019yox,Zha:2020cst,Zha:2017jch}. This further inspires experimentalists and theorists to explore the existence of coherent photon-photon interactions in HHIC, another important subject that has also been exclusively studied in UPC. Building on this line of reasoning, STAR and ATLAS also identified significant excesses in dilepton production at very low $p_T$ in HHIC~\cite{STAR:2018ldd,ATLAS:2022yad}, and these have been effectively elucidated by introducing the coherent photon-photon production mechanism~\cite{Klein:2018cjh,Klusek-Gawenda:2018zfz,Zha:2018ywo}. These advancements in both experimental and theoretical realms have facilitated the integration of physics related to UPC into HHIC, accompanied by the presence of vigorous strong interactions within the overlap region.

The presence of photoproduction in hadronic collisions has the potential to influence our comprehension of QGP deduced from prior measurements of $J/\psi$ suppression, as these measurements did not account for the contribution from photoproduction. The degree of suppression for $J/\psi$ is quantified by the nuclear modification factor ($R_{AA}$), which establishes the connection between the $J/\psi$ yield observed in heavy-ion collisions and that in proton-proton ($p$+$p$) collisions. Therefore, in order to elucidate the impact of photoproduction, it is imperative to investigate the photoproduction contribution concurrently in both hadronic $p$+$p$ and A+A collisions. Fortunately, our prior study~\cite{Cao:2018hvy} demonstrated that the contribution from photoproduction has a negligible effect on the $p$+$p$ baseline. In this paper, we proceed to investigate its impact in hadronic A+A collisions, extracting the authentic signals of $J/\psi$ suppression. We calculate the coherent and incoherent photoproduction of $J/\psi$ particles at both forward and mid-rapidity in Au+Au and Cu+Cu collisions at $\sqrt{s_{\textrm{NN}}}=200$ GeV. We present the $J/\psi$ nuclear modification factor ($R_{AA}$) contributed by photoproduction as a function of $p_T$ and number of participating nucleons ($N_{part}$) in Au+Au and Cu+Cu collisions. Furthermore, we deduct the contribution from photoproduction from the overall yield, resulting in the corrected $R_{AA}$. This corrected factor serves as a more precise indicator for probing the properties of QGP.

\section{METHODOLOGY}
Following Ref. \cite{Zha:2020cst}, the photoproduction probability of $J/\psi$ between the two colliding nuclei in momentum space can be expressed as follows:
\begin{equation} 
\frac{d^2P}{dp_{x}dp_{y}}=\frac{1}{2\pi}|\int  {(A_1( \vec{x}_{\perp})+A_2({\vec{x}_{\perp}}))e^{i p_{\perp}\cdot {\vec{x}_{\perp}}}\,d^2\vec{x}_{\perp}}|^2, \label{eq:one}
\end{equation}

where, $\vec{x}_{\perp}$ and $p_{\perp}$ denote the two-dimensional spatial coordinate and momentum vector, respectively. The $A_1(\vec{x}_{\perp})$ and $A_2(\vec{x}_{\perp})$ represent the amplitudes governing the production of $J/\psi$ via photon-Pomeron fusion on the two respective nuclei. The production amplitudes $A_{1,2}(\vec{x}{\perp})$ are determined by the spatial photon flux and the $\gamma A$ scattering amplitude $\Gamma_{\gamma A\rightarrow VA}$. The photon flux is described by the equivalent photon approximation~\cite{Krauss:1997vr,Klein:1999qj}:

\begin{equation*}
n(\omega_{\gamma},\vec{x}_{\perp})= \frac{4Z^2\alpha}{\omega_{\gamma}} \Big| 
\int \frac{d^2\vec{k}_{\gamma \perp}}{(2\pi)^2}
  \vec{k}_{\gamma} \frac{F_{\gamma}(\vec{k}_{\gamma})}{|\vec{k}_{\gamma}|^2}e^{i\vec{x}_{\perp} \cdot \vec{k}_{\perp}}\Big|^2 , 
\end{equation*}

\begin{equation}
 \vec{k}_{\gamma}=(\vec{k}_{\gamma\perp},\frac{\omega_{\gamma}}{\gamma_c})\; ,\quad\quad \omega_{\gamma}=\frac{1}{2}M_{J/\psi}e^{\pm y}\; ,  \label{eq:two}
\end{equation}
where, $\vec{x}_{\perp}$ and $\vec{k}_{\gamma\perp}$ represent the 2-dimensional position and momentum vectors of the photon perpendicular to the beam direction. $Z$ denotes the nuclear charge, $\alpha$ stands for the electromagnetic coupling constant, $\gamma_c$ corresponds to the Lorentz factor of the photon-emitting nucleus, $M_{J/\psi}$ and $y$ respectively indicate the mass and rapidity of $J/\psi$. $F_{\gamma}(\vec{k}_{\gamma})$ is the nuclear electromagnetic form factor, determined via Fourier transformation of the nucleus's charge density. The charge density of a nucleus is described by the Woods-Saxon distribution:
\begin{equation}
\rho_A(r)=\frac{\rho^0}{1+exp[(r-R_{WS})/d]} \; ,
\end{equation}
where, $r$ represents the distance from the nucleus's center, and $R_{WS}$ and $d$ are respectively the radius and skin depth of the nucleus, which can be determined from fits to electron-scattering data \cite{1977Nuclear}. The normalization factor is denoted as $\rho^0$. The parameters of Au and Cu employed in calculations are provided in Table \ref{table:1}.
\begin{table}[h]
\caption{ Woods-Saxon parameters of Au and Cu.}
\begin{center}
\begin{tabular}{  m{1.5cm} m{1.5cm} m{1.5cm} m{1.5cm} m{1.5cm} }
\hline
 & A & Z & ${R_{WS}/fm}$ & $d/fm$ \\
\hline
\quad Au & 197 & 79 & \quad6.38 & 0.545 \\
\hline
\quad Cu & 63 & 29 & \quad4.20 & 0.596 \\
\hline
\end{tabular}
\end{center}
\label{table:1}
\end{table}

The scattering amplitude $\Gamma_{\gamma A \rightarrow VA}$, incorporating shadowing effects, can be estimated using the quantum Glauber \cite{Miller:2007ri} combined with the vector meson dominance (VMD) approach \cite{Bauer:1977iq}:
\begin{equation}
\Gamma_{\gamma A \rightarrow VA}(\vec{x}_{\perp})=\frac{f_{\gamma N \rightarrow VN}(0)}{\sigma_{VN}}2\Bigr[1-exp\Bigl(-\frac{\sigma_{VN}}{2}T^{'}(\vec{x}_{\perp})\Bigl)\Bigr] \;  ,
\end{equation}
where, $f_{\gamma N \rightarrow VN}(0)$ represents the forward-scattering amplitude for the process $\gamma + N \rightarrow V + N$, where the number 0 in parenthesis signifies zero momentum transfer from photon to nucleus. The cross section $\sigma_{VN}$ corresponds to the vector meson-nucleon scattering. The parameter $f_{\gamma N \rightarrow VN}(0)$ is determined through a global fit to experimental data \cite{Klein:2016yzr}. Due to the coherent effect in the longitudinal ($z$) direction, an adjustment is necessary for the thickness function:
\begin{equation*}
T^{'}(\vec{x}_{\perp})=\int _{-\infty}^{\infty} \rho_A\biggl(\sqrt{\vec{x}_{\perp}^{2}+z^2}\biggl)e^{iq_{L}z} \, dz\; ,
\end{equation*}
\begin{equation}
q_L=\frac{M_V\cdot e^{\mp y}}{2\gamma_c}\; ,
\end{equation}
where $q_L$ signifies the longitudinal momentum transfer necessary for producing an actual vector meson. The total cross section for vector meson-nucleon scattering ($VN$) can be calculated using the optical theorem and the VMD relation:
\begin{equation}
\sigma_{VN}=\frac{f_V}{4\sqrt{\alpha}C}f_{\gamma N \rightarrow VN}\; ,
\end{equation}
where $f_V$ represents the coupling constant between the vector meson and photon ($V$-photon coupling), and $C$ is a correction factor accounting for the nondiagonal coupling through higher mass vector mesons. The production amplitudes can be then expressed as follows:
\begin{equation*}
A_1(\vec{x}_{\perp})=\Gamma_{\gamma A_1 \rightarrow VN}\sqrt{n_2(\omega_{\gamma},\vec{x}_{\perp})},
\end{equation*}
\begin{equation}
A_2(\vec{x}_{\perp})=\Gamma_{\gamma A_2 \rightarrow VN}\sqrt{n_1(\omega_{\gamma},\vec{x}_{\perp})},
\end{equation}
where, $A_1(\vec{x}{\perp})$ indicates that nucleus 2 emits the photons, while $A_2(\vec{x}{\perp})$ indicates that nucleus 1 emits the photons.

According to our previous study~\cite{Zha:2017jch}, the presence of intense strong interactions should preclude photoproduction within the nuclear overlap region. As a result of this consideration, the amplitude undergoes modification:

\begin{equation}
A_1^{'}(\vec{x}_{\perp}) =A_1(\vec{x}_{\perp})\Bigl(1-T_2(\vec{x}_{\perp})\cdot \sigma_{inel}\Bigl)^A \; ,
\label{eight}
\end{equation}
where $\sigma_{inel}$ represents the inelastic cross section for proton-proton collisions, taking the value $\sigma_{inel}$ = 42 mb. $A$ denotes the atomic number of the nucleus. $[1-T_2(\vec{x}{\perp})\cdot \sigma_{inel}]^A$ quantifies the probability of no hadronic interaction occurring. Ultimately, the yield of $J/\psi$ can be determined by substituting Eq.~\ref{eight} into Eq.~\ref{eq.one}. The cross section $d\sigma/dy$ in a specific centrality is calculated by integrating $dN/dy$ over the corresponding impact parameter ($b$) range:
\begin{equation}
d\sigma/dy(J/\psi)=\int _{b_{min}}^{b_{max}} 2\pi b dN/dy(J/\psi,b) \, db \; .
\end{equation}

In addition to coherent photoproduction, incoherent photoproduction also contributes to the total $J/\psi$ yield in hadronic heavy-ion collisions. The incoherent cross section $\sigma_{\gamma, A \rightarrow J/\psi, A^\prime}$ can be approximated by scaling the cross section $\sigma_{\gamma,p \rightarrow J/\psi,p}$ via the application of the Glauber + VMD approach. Where, $A^\prime$ signifies the nucleus having disintegrated into other product fragments. The cross section can be formulated as follows:  

\begin{equation*}
{\sigma_{\gamma A \rightarrow J/\psi A^\prime}} = \sigma_{\gamma p \rightarrow J/\psi p}  
 \int T(\vec{x}_{\perp})e^{(-1/2)\sigma_{VN}^{in} T(\vec{x}_{\perp})}  \, d^2\vec{x}_{\perp},
\end{equation*}
\begin{equation}
\sigma_{VN}^{in}=\sigma_{VN}-\sigma_{VN}^2/(16\pi B_V)\; ,
\end{equation}
where $T(\vec{x}{\perp})$ represents the thickness function of the nucleus, $\sigma_{VN}^{in}$ corresponds to the inelastic vector meson-nucleon cross section, and $B_V$ signifies the slope of the $t$ dependence in the $\gamma p \rightarrow Vp$ scattering process.

\section{Results}
\begin{figure}[ht]
\includegraphics[width=0.95\columnwidth]{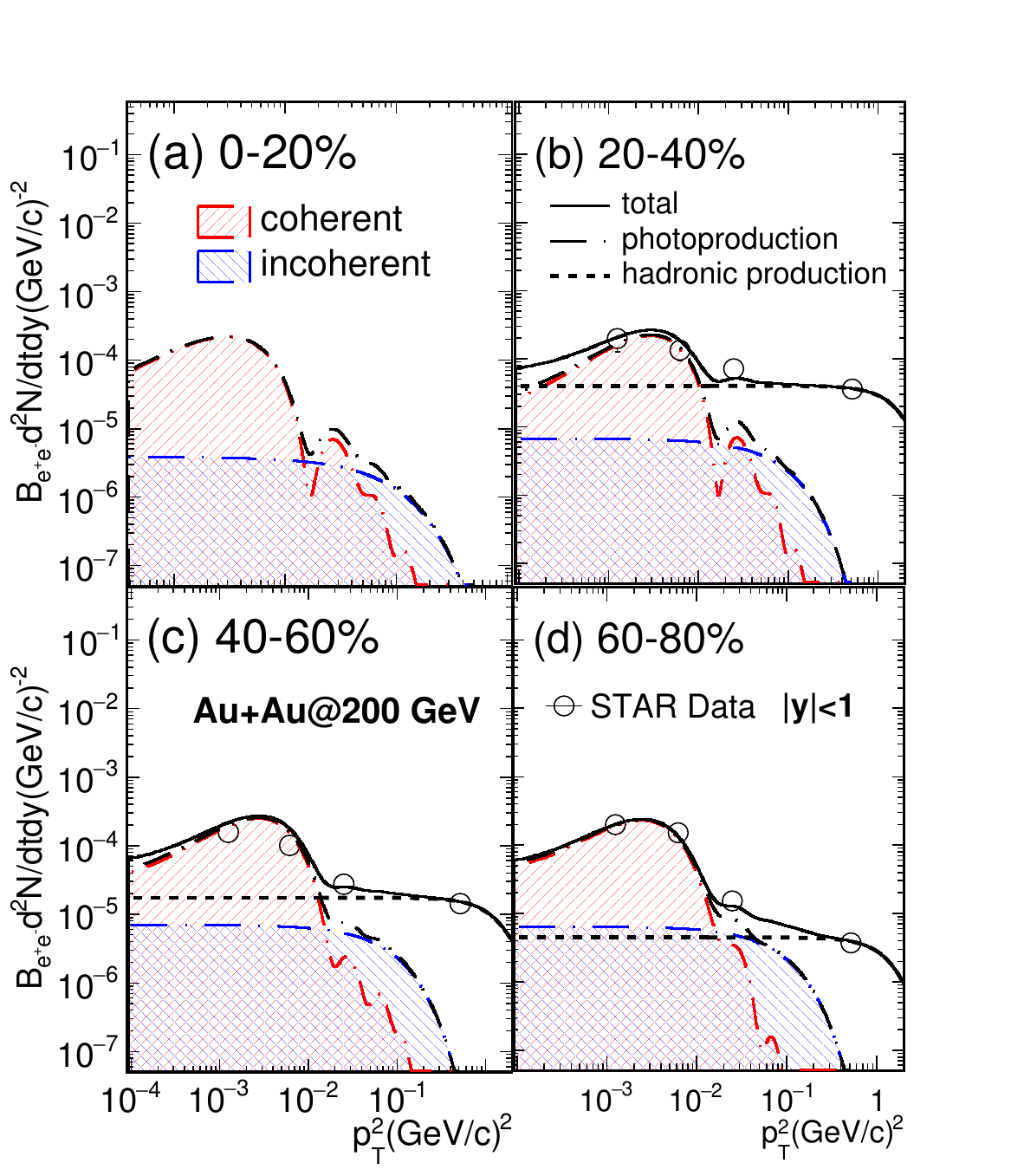}
\caption{Distributions of photon-induced $J/\psi$ invariant yield as a function of $-t\sim p_T^2$ for mid-rapidity($|y|<1$) in Au+Au collisions at $\sqrt{s_{\textrm{NN}}}=200$ GeV with centralities of (a)0-20$\%$, (b)20-40$\%$, (c)40-60$\%$, (d)60-80$\%$. The open circles represent STAR data taken from \cite{STAR:2019yox}.}
\label{fig:mesh1}
\end{figure}

Figure \ref{fig:mesh1} illustrates the $J/\psi$ $d^2N/dtdy$ distributions resulting from photoproduction at mid-rapidity ($|y|<1$) in different centralities Au+Au collisions. The Mandelstam variable $t$ is approximately equivalent to -$p_T^2$ at the top energy of RHIC. The red and blue shaded regions correspond respectively to the contributions from coherent and incoherent photoproduction. The expected hadronic contributions, depicted as dotted lines, are obtained through fits to the STAR experimental data points~\cite{STAR:2019yox} for $p_T>0.2$ GeV/$c$, following the equation:
\begin{equation}
\frac{d^2N}{2\pi\,p_Tdp_Tdy}=\frac{a}{(1+b^2p_T^2)^n},
\end{equation}
where $a$, $b$, $n$ are free parameters. The black solid lines portray the combined contribution of photoproduction and hadronic production. It describes the experimental data in the entire $p_T$ range. The $J/\psi$ yields from photoproduction decrease rapidly with increasing $p_T$, and they consistently remain considerably smaller than the hadronic production for $p_T>0.2$ GeV/$c$. Upon scrutinizing the four panels in Figure \ref{fig:mesh1}, it becomes apparent that photoproduction exhibits a more pronounced impact in more peripheral collisions.

\begin{figure}[ht]
\includegraphics[width=0.95\columnwidth]{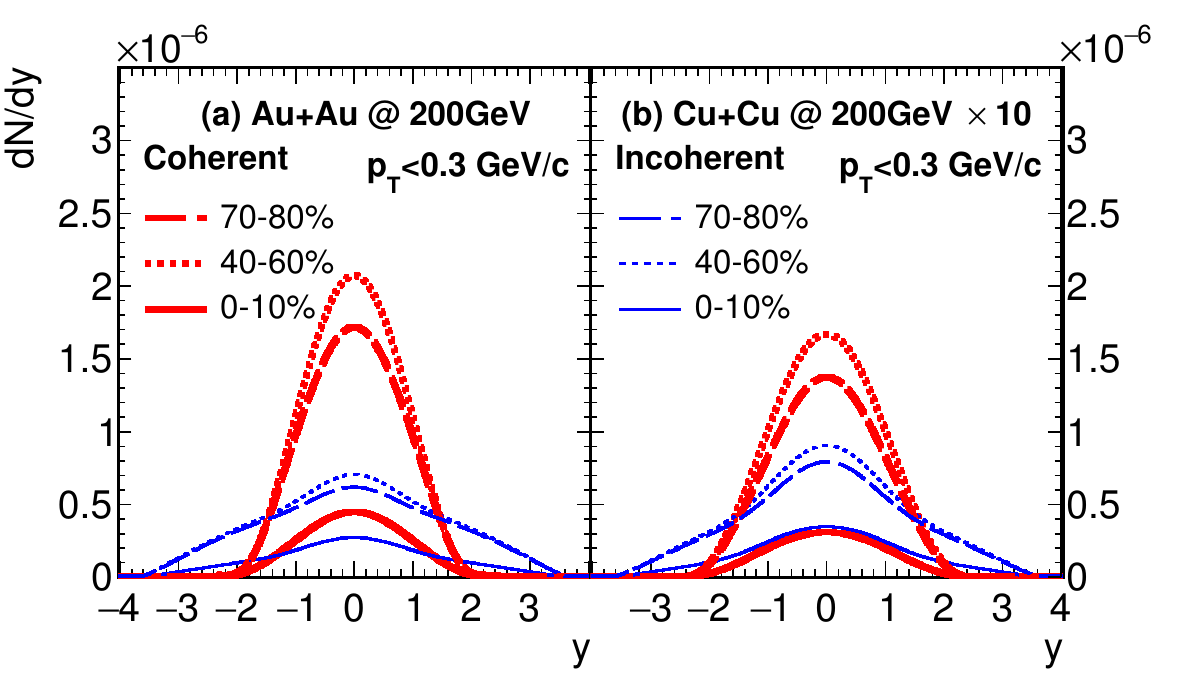}
\caption{$dN/dy$ distributions of $J/\psi$ from coherent and incoherent photoproduction in (a) Au+Au and (b) Cu+Cu collisions at $\sqrt{s_{\textrm{NN}}} = 200$ GeV.}
\label{fig:mesh2}
\end{figure}

Figure~\ref{fig:mesh2} shows the J/$\psi$ $dN/dy$ distributions of coherent and incoherent photoproduction in different centralities (a) Au+Au and (b) Cu+Cu collisions. The thick red lines signify coherent photoproduction, while the blue lines denote incoherent photoproduction. The rapidity distributions in Cu+Cu collisions have been scaled by a factor of 10 for clarity. The coherent and incoherent photoproduction appears diminished in central collisions due to destructive interference and external influences from the nuclear overlap region. The increase in cross section from 40-60$\%$ centrality to 70-80$\%$ centrality is attributed to an augmented photon flux originating from nuclei with smaller impact parameters.  At forward rapidity, the contraction of the coherent length results in a decrease in coherent photoproduction. This phenomenon, however, has no impact on incoherent photoproduction, thereby rendering incoherent photoproduction more prominent at forward rapidity. 

\begin{figure}[t] 
\includegraphics[width=1.0\columnwidth]{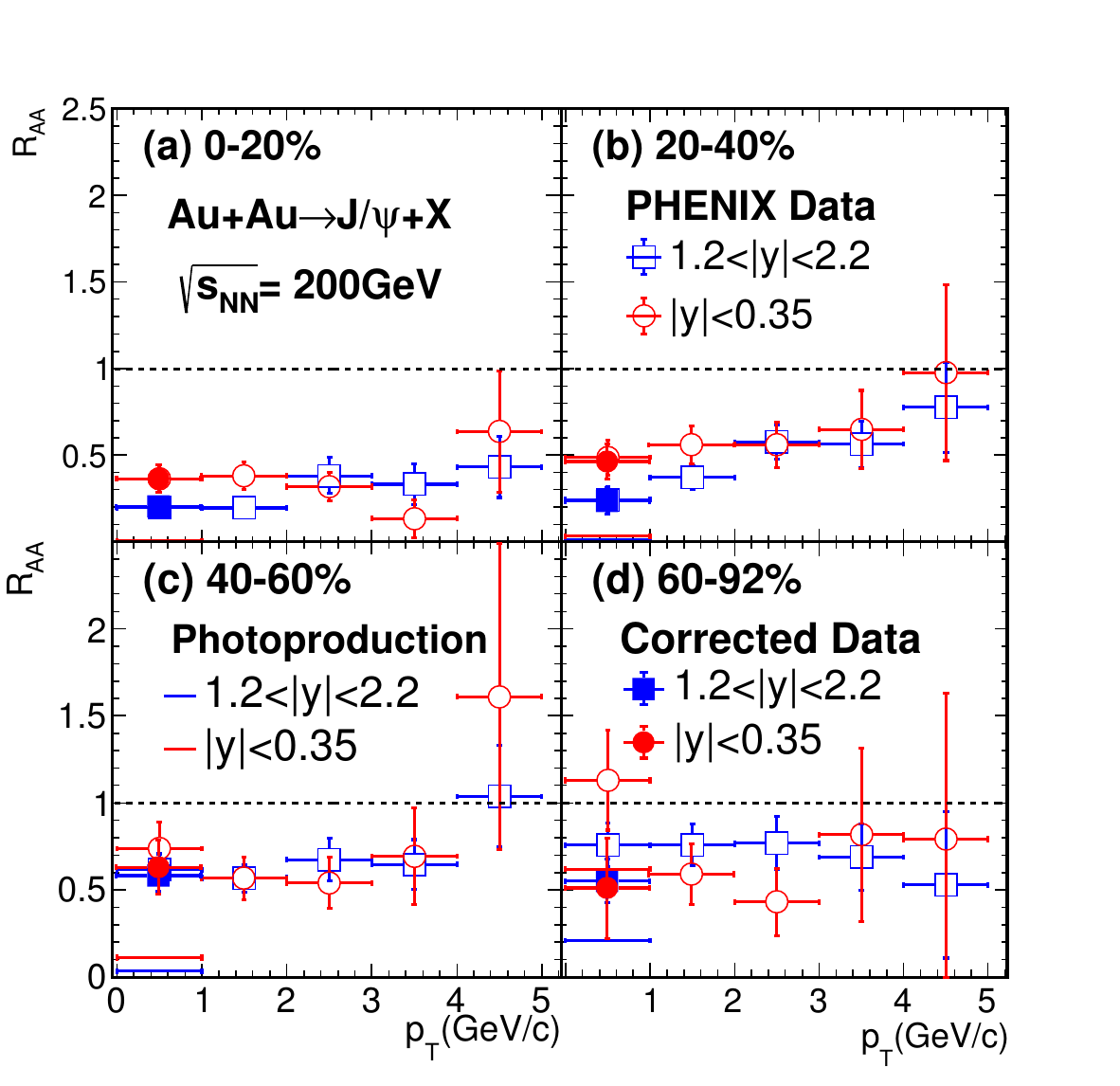} 
\caption{$J/\psi$ $ R_{AA}$ as a function of $p_T$ at both forward and mid-rapidity in different centralities: (a) 0-20$\%$, (b) 20-40$\%$, (c) 40-60$\%$, and (d) 60-92$\%$, Au+Au collisions at $\sqrt{s_{\textrm{NN}}}=200$ GeV.}
\label{fig:mesh3}
\end{figure}
\noindent

Figure \ref{fig:mesh3} shows $J/\psi$ $R_{AA}$ as a function of $p_T$ at both forward and mid-rapidity in different centralities Au+Au collisions. The red open circles represent the $R_{AA}$ data at mid-rapidity, while the blue open squares correspond

\begin{figure}[ht] 
\includegraphics[width=1.0\columnwidth]{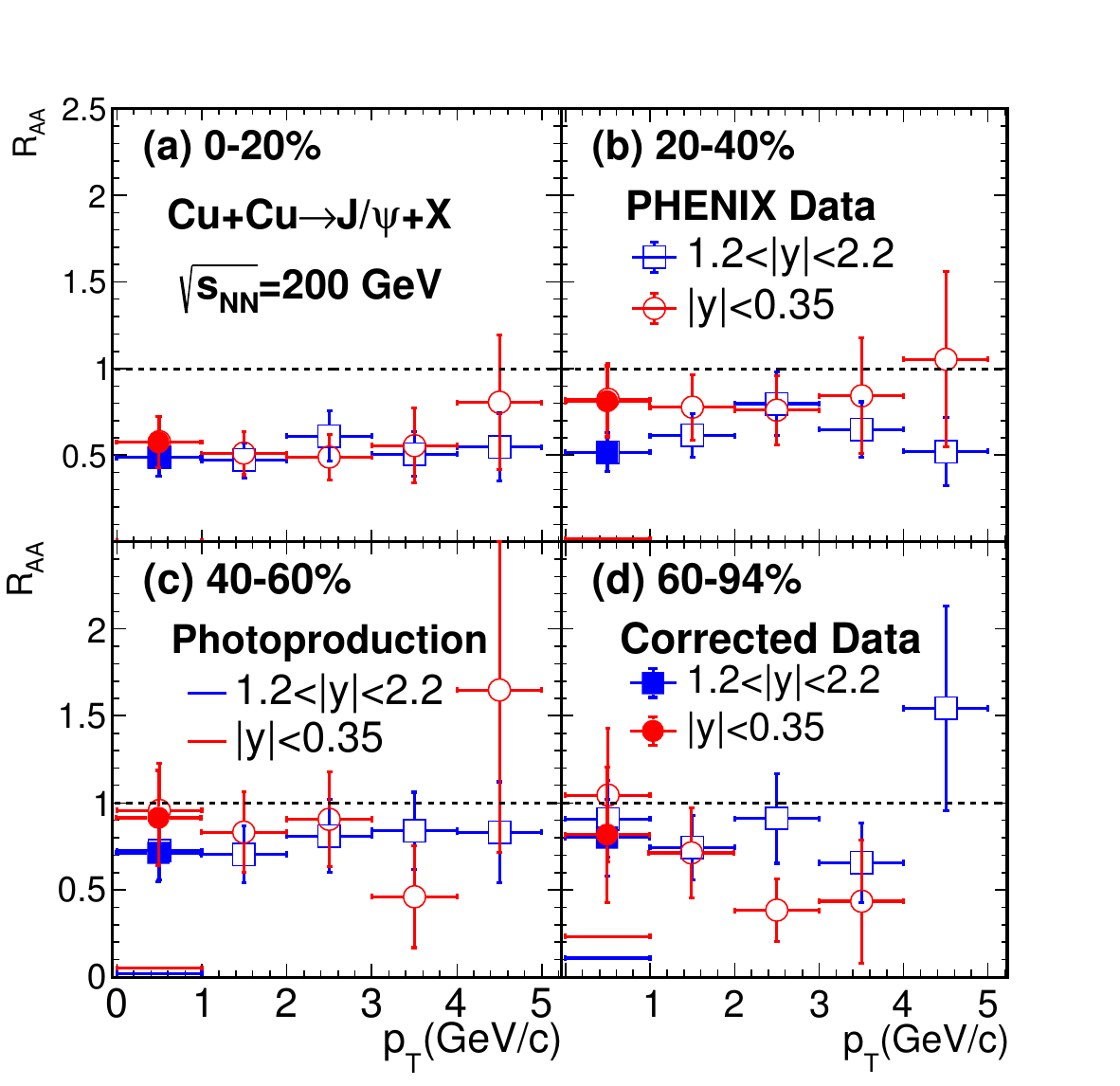}
\caption{$J/\psi$ $R_{AA}$ as a function of $p_T$ at both forward and mid-rapidity in different centralities: (a) $0-20\%$, (b) $20-40\%$, (c) $40-60\%$, and (d) $60-94\%$, Cu+Cu collisions at $\sqrt{s_{\textrm{NN}}}=200$ GeV.}
\label{fig:mesh4}
\end{figure}

\noindent
to the $R_{AA}$ data at forward rapidity measured by PHENIX collaboration \cite{PHENIX:2011img,PHENIX:2008jgc}. The calculation of $R_{AA}$ is performed by dividing the $J/\psi$ yield in A+A collisions by the scaled yield in $p+p$ collisions using the formula:
\begin{equation}
R_{AA}=\frac{\sigma_{inel}}{\langle N_{coll} \rangle}\frac{d^2N_{AA}/dydp_T}{d^2\sigma_{pp}/dydp_T}\; ,
\end{equation}
where, $d^2N_{AA}/dydp_T$ represents the $J/\psi$ yield in A+A collisions, and $d^2\sigma_{pp}/dydp_T$ denotes the $J/\psi$ cross section in $p$+$p$ collisions. The values of $\langle N_{coll} \rangle$, $\langle N_{part} \rangle$, and $\langle b \rangle$ along with their associated systematic uncertainties for each centrality are sourced from \cite{PHENIX:2011img,PHENIX:2014kia}. For showcasing the influence of photoproduction, horizontal lines are depicted, indicating the photoproduction contribution for the low $p_T$ range ($p_T<1$ GeV/$c$) in each centrality. Since photoproduction becomes significantly negligible compared to hadronic production when $p_T>1$ GeV/$c$, it is not shown in the figure for those regions. The solid squares and circles represent the corrected $R_{AA}$ values after accounting for the subtraction of photoproduction contributions. The error bars encapsulate both statistical and systematic uncertainties.

Figure \ref{fig:mesh4} depicts $J/\psi$ $R_{AA}$ as a function of $p_T$ at both forward and mid-rapidity in different centralities Cu+Cu collisions.

\begin{table*}[!htbp]
\caption{The yield and $R_{AA}$ values for coherent and incoherent photoproduction at mid and forward rapidity in different centralities for both Au+Au and Cu+Cu collisions at 200 GeV.}
\begin{ruledtabular}
\begin{tabular}{ccccc}
  & Centrality ($\%$) & $Br_{e^+e^-}\frac{dN}{dy}$(coherent) & $Br_{e^+e^-}$ $\frac{dN}{dy}$(incoherent)  & $R_{AA}$ \\ \hline 
\multirow{4}{8em}{Au+Au $\newline|y|<0.35\newline p_T<1$ GeV/$c$}
& 0-20  & 8.5$\times10^{-7}$  & 3.9$\times10^{-7}$  &  0.006  \\ 
& 20-40 & 1.9$\times10^{-6}$  & 6.7$\times10^{-7}$  &  0.03  \\
& 40-60 & 2.0$\times10^{-6}$  & 7.0$\times10^{-7}$  &  0.11  \\
& 60-92 & 1.6$\times10^{-6}$  & 6.0$\times10^{-7}$  &  0.62  \\ \hline
\multirow{4}{8em}{Au+Au $\newline1.2<|y|<2.2\newline p_T<1$ GeV/$c$} 
& 0-20  & 1.4$\times10^{-7}$  & 1.9$\times10^{-7}$  &  0.002 \\
& 20-40 & 2.6$\times10^{-7}$  & 3.5$\times10^{-7}$  &  0.01  \\
& 40-60 & 2.8$\times10^{-7}$  & 3.9$\times10^{-7}$  &  0.04  \\
& 60-92 & 2.5$\times10^{-7}$  & 3.6$\times10^{-7}$  &  0.21  \\ \hline 
\multirow{4}{8em}{Cu+Cu $\newline |y|<0.35\newline p_T<1$ GeV/$c$}
& 0-20  & 6.1$\times10^{-8}$  & 4.8$\times10^{-8}$  &  0.003 \\ 
& 20-40 & 1.5$\times10^{-7}$  & 8.4$\times10^{-8}$  &  0.02  \\
& 40-60 & 1.7$\times10^{-7}$  & 9.0$\times10^{-8}$  &  0.05  \\
& 60-94 & 1.2$\times10^{-7}$  & 7.3$\times10^{-8}$  &  0.23  \\ \hline
\multirow{4}{8em}{Cu+Cu $\newline1.2<|y|<2.2\newline p_T<1$ GeV/$c$} 
& 0-20  & 1.5$\times10^{-8}$  & 2.0$\times10^{-8}$  &  0.001  \\
& 20-40 & 3.5$\times10^{-8}$  & 3.7$\times10^{-8}$  &  0.006 \\
& 40-60 & 3.6$\times10^{-8}$  & 4.0$\times10^{-8}$  &  0.02  \\
& 60-94 & 2.8$\times10^{-8}$  & 3.5$\times10^{-8}$  &  0.11  \\ 
\end{tabular}
\end{ruledtabular}
\label{table:3}
\end{table*}
\vspace{6cm}

The coherent and incoherent photon-induced $J/\psi$ yields in Au+Au and Cu+Cu collisions are listed in Table~\ref{table:3}. The greater significance of coherent and incoherent photoproduction in Au+Au collisions, compared to Cu+Cu collisions, can be attributed to the larger photon flux, which scales with the square of the nuclear charge. Moreover, the photon-induced process demonstrates less sensitivity to changes in centralities when compared to hadronic production, which is much smaller in peripheral collisions because of the much less $N_{coll}$. Interestingly, the yields of photoproduction even show an increase in more peripheral collisions. This phenomenon results in a more pronounced contribution of photoproduction to $J/\psi$ $R_{AA}$ in peripheral collisions. Specifically, the $R_{AA}$ contributed by photoproduction can reach approximately 0.6 at mid-rapidity ($|y|<0.35$) in the 60-92$\%$ Au+Au collisions and around 0.2 at mid-rapidity ($|y|<0.35$) for the 60-94$\%$ Cu+Cu collisions at $p_T$ range of 0-1 GeV/$c$. Consequently, it's important not to disregard the contribution of photoproduction in peripheral collisions at low $p_T$ region. Referring to Eq.~\ref{eq:two}, it can be deduced that the photon flux inversely correlates with the energy of the photon, and since the photon's energy is proportional to the rapidity of $J/\psi$, this leads to a decrease in the photon flux converted to $J/\psi$ with increasing $J/\psi$ rapidity. This insight further underscores the greater contribution of photoproduction in mid-rapidity. This observation aligns with the calculated results, where the photon-induced $R_{AA}$ at mid-rapidity are approximately three times those at forward rapidity in each centrality Au+Au (Cu+Cu) collisions.

\begin{figure}[ht]
\includegraphics[width=0.95\columnwidth]{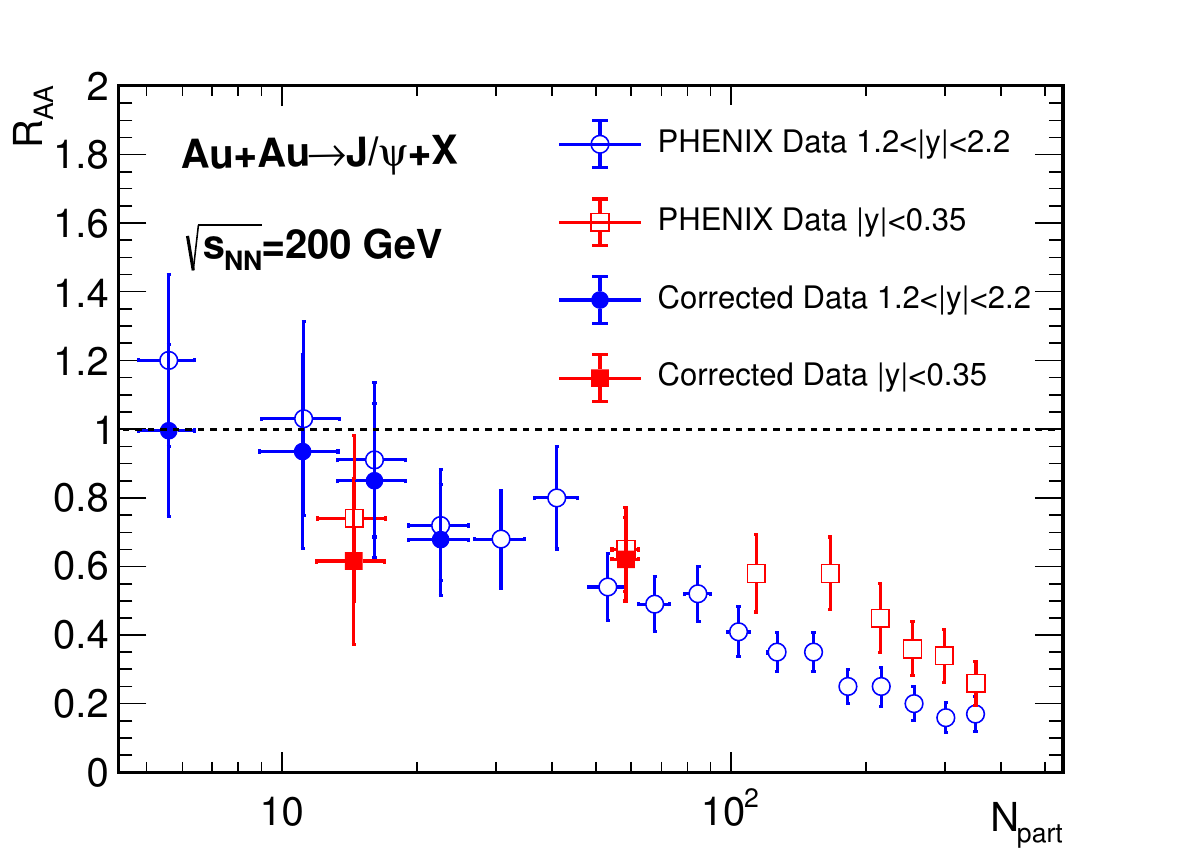}
\caption{The $J/\psi$ $p_T$-integrated $R_{AA}$ as a function of $N_{part}$ in Au+Au collisions at $\sqrt{s_{\textrm{NN}}}=200$ GeV.}
\label{fig:mesh5}
\end{figure}

\begin{figure}[ht]
\includegraphics[width=0.95\columnwidth]{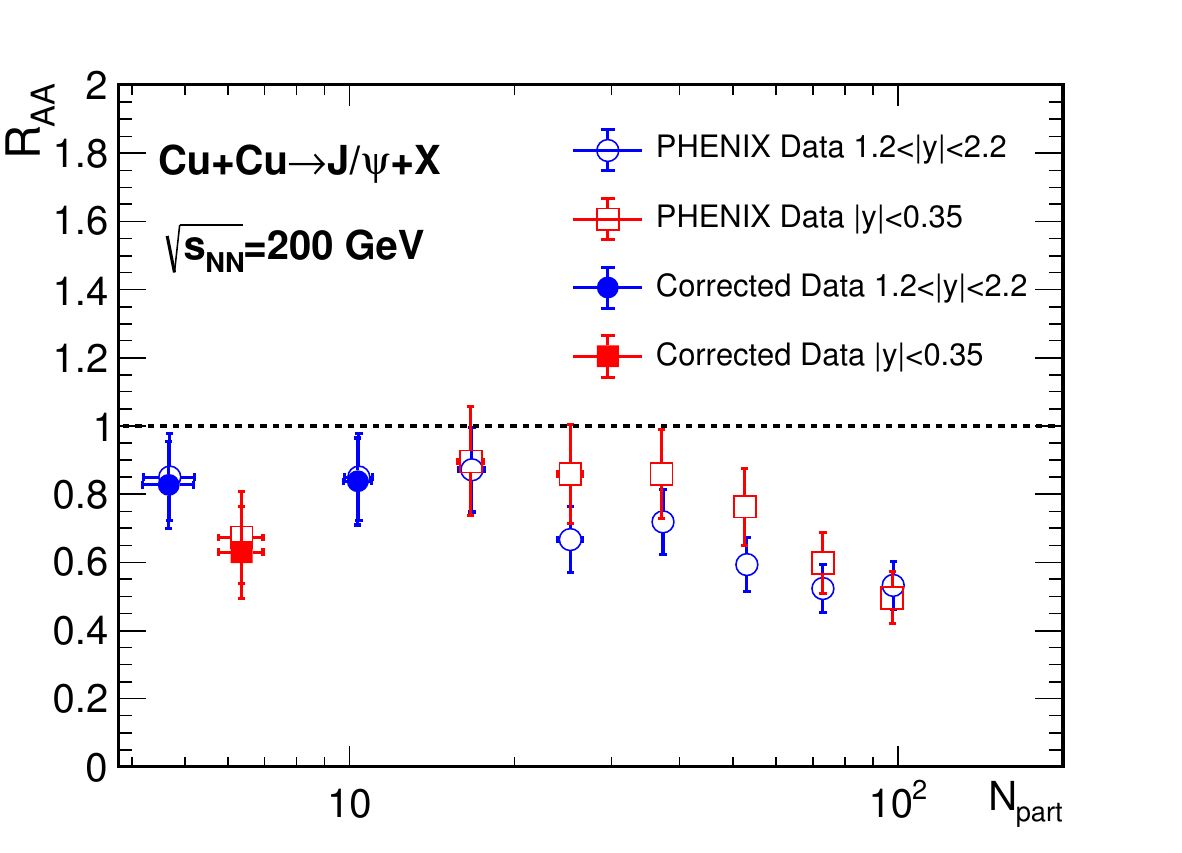}
\caption{The $J/\psi$ $p_T$-integrated $R_{AA}$ as a function of $N_{part}$ in Cu+Cu collisions at $\sqrt{s_{\textrm{NN}}}=200$ GeV.}
\label{fig:mesh6}
\end{figure}

Figure~\ref{fig:mesh5} presents the $p_T$-integrated $J/\psi$ $R_{AA}$ as a function of $N_{part}$ in Au+Au collisions. The experimental data from PHENIX are depicted using red open squares for mid-rapidity and blue open circles for forward rapidity. The corrected data are further displayed with red solid squares for mid-rapidity and blue solid circles for forward rapidity. Notably, photoproduction is primarily significant in very peripheral collisions and negligible in central collisions. Thus, the presentation of corrected $R_{AA}$ focuses on small $N_{part}$ values. Given the propensity of photoproduction to be more prominent at low $p_T$, the corrections applied to the $p_T$-integrated $R_{AA}$ values are relatively minor. The $R_{AA}$ contributions from photoproduction can reach values of 0.2 for forward rapidity in the 80-92$\%$ centrality and 0.07 for mid-rapidity in the 60-92$\%$ centrality. Figure \ref{fig:mesh6} displays the $p_T$-integrated $J/\psi$ $R_{AA}$ as a function of $N_{part}$ in Cu+Cu collisions. Notably, the corrections in Cu+Cu collisions are not as pronounced as in Au+Au collisions.

\section{Summary}

In this study, we conducted comprehensive calculations of coherent and incoherent photon-induced $J/\psi$ production at $\sqrt{s_{\textrm{NN}}}$=200 GeV in both Au+Au and Cu+Cu collisions. Our analysis encompassed the examination of photon-induced $J/\psi$ yields and their corresponding contribution to the nuclear modification factors ($R_{AA}$) in Au+Au and Cu+Cu collision systems. Notably, the contributions stemming from photoproduction prove substantial in very peripheral collisions and display a heightened significance at mid-rapidity. Particularly intriguing, $R_{AA}$ contributions attributed to photoproduction can extend up to approximately 0.6 at low $p_T$ ($0<p_T<1$ GeV/$c$) in the mid-rapidity region ($|y|<0.35$) in 60-92$\%$ Au+Au collisions. In contrast, the corrections observed in Cu+Cu collisions are less pronounced. Moreover, we showcased the  $J/\psi$ $p_T$-integrated $R_{AA}$ as a function of $N_{part}$ in both Au+Au and Cu+Cu collisions. Notably, the correction is only sizable as $N_{part}<20$ in Au+Au and Cu+Cu collisions. By incorporating the effects of photoproduction, the $J/\psi$ suppression measurements harmonize more effectively with the overarching physical framework governing the interplay between cold and hot medium effects in heavy-ion collisions. This allows us to extract the properties of QGP more precisely through $J/\psi$ measurements.
\vspace{5cm}
\nocite{*}

\bibliography{apssamp}% Produces the bibliography via BibTeX.
%\end{CJK}
\end{document}